\documentclass[12pt]{article}

\usepackage{scicite}
\usepackage{times}
\usepackage{graphicx}
\usepackage{bm}
\usepackage{bbm}
\usepackage{mathabx}

\newenvironment{sciabstract}{%
\begin{quote} \bf}
{\end{quote}}

\newcounter{lastnote}

\newcommand{\op}[1]{\hat{#1}}				
\newcommand{\vect}[1]{{\bm{#1}}}			
\newcommand{\avg}[1]{\langle#1\rangle}		
\newcommand{\bra}[1]{\langle#1|}			
\newcommand{\ket}[1]{|#1\rangle}			
\newcommand{\upket}{|\!\!\uparrow\rangle}	
\newcommand{\upbra}{\langle\uparrow\!\!|}	
\newcommand{\eg}{\emph{e.g.}}			
\newcommand{\ie}{\emph{i.e.}}				
\newcommand{\dagg}{^{\dagger}}			
\newcommand{\si}[1]{_{\mathrm{#1}}}		
\newcommand{\dd}{\mathrm{d}}			
\newcommand{\one}{\mathbbm{1}}			

\makeatletter
\newcommand{\underset}[2]{\ensuremath{\mathop{\kern\z@\scalebox{1.15}{\mbox{\ensuremath{#2}}}}\limits_{\scalebox{0.55}{\mbox{\ensuremath{#1}}}}}}
\makeatother

\title{Bell Correlations in a Bose-Einstein Condensate}

\author
{Roman Schmied,$^{1\ast}$
Jean-Daniel Bancal,$^{2,4\ast}$
Baptiste Allard,$^{1\ast}$
Matteo Fadel,$^{1}$\\
Valerio Scarani,$^{2,3}$
Philipp Treutlein,$^{1\dagg}$
Nicolas Sangouard$^{4\dagg}$\\
\\
\normalsize{$^{1}$Quantum Atom Optics Lab, Department of Physics, University of Basel,}\\
\normalsize{Klingelbergstrasse 82, 4056 Basel, Switzerland,}\\
\normalsize{$^{2}$Centre for Quantum Technologies, National University of Singapore,}\\
\normalsize{3 Science Drive 2, Singapore 117543}\\
\normalsize{$^{3}$Department of Physics, National University of Singapore,}\\
\normalsize{2 Science Drive 3, Singapore 117542}\\
\normalsize{$^{4}$Quantum Optics Theory Group, Department of Physics, University of Basel,}\\
\normalsize{Klingelbergstrasse 82, 4056 Basel, Switzerland,}\\
\\
\normalsize{$^{\ast}$These authors contributed equally to this work.}\\
\normalsize{$^{\dagg}$Corresponding authors. E-mail: philipp.treutlein@unibas.ch, nicolas.sangouard@unibas.ch}
}

\date{\today}

\begin{document}

\maketitle

\begin{sciabstract}
Characterizing many-body systems through the quantum correlations between their constituent particles is a major goal of quantum physics. Although entanglement is routinely observed in many systems, we report here the detection of stronger correlations -- Bell correlations --  between the spins of about 480 atoms in a Bose-Einstein condensate. We derive a Bell correlation witness from a many-particle Bell inequality involving only one- and two-body correlation functions. Our measurement on a spin-squeezed state exceeds the threshold for Bell correlations by 3.8 standard deviations. Our work shows that the strongest possible non-classical correlations are experimentally accessible in many-body systems, and that they can be revealed by collective measurements.
\end{sciabstract}

\maketitle 

Parts of a composite quantum system can share correlations that are stronger than any classical theory allows~\cite{BrunnerRMP2014}. These so-called Bell correlations represent the most profound departure of quantum from classical physics and can be confirmed experimentally by showing that a Bell inequality is violated by the system. The existence of Bell correlations at space-like separations refutes local causality~\cite{Bell1990}; thus, Bell correlations are also called nonlocal correlations. Moreover, they are a key resource for quantum technologies such as quantum key distribution and certified randomness generation~\cite{ScaraniAPS2012}. Bell correlations have so far been detected between up to fourteen ions~\cite{LanyonPRL2014}, four photons~\cite{EiblPRL2003,ZhaoPRL2003}, two neutral atoms~\cite{HofmannSCIENCE2012}, two solid-state spin qubits~\cite{PfaffNATUREPHYSICS2013}, and two Josephson phase qubits~\cite{AnsmannNATURE2009}. Even though multi-partite Bell inequalities are known~\cite{DrummondPRL1983,SvetlichnyPRD1987,ZukowskiPRL2002,BrunnerRMP2014}, the detection of Bell correlations in larger systems is challenging.

A central challenge in quantum many-body physics is to connect the global properties of a system to the underlying quantum correlations between the constituent particles~\cite{AmicoRMP2008,BlochRMP2009}. For example, recent experiments in quantum metrology have shown that spin-squeezed states of atomic ensembles can enhance the precision of interferometric measurements beyond classical limits~\cite{GrossNATURE2010,LerouxPRL2010,LouchetChauvetNJP2010,OckeloenPRL2013}. This enhancement requires  entanglement between atoms in the ensemble, which can be revealed by measuring an entanglement witness that involves only collective measurements on the entire system~\cite{SoerensenNATURE2001,SoerensenPRL2001,GrossNATURE2010,RiedelNATURE2010,HyllusPRA2012}. The role of Bell correlations in many-body systems, on the other hand, is largely unknown. Whereas all Bell-correlated states are entangled, the converse is not true~\cite{BrunnerRMP2014}. In recent theoretical work, a family of Bell inequalities was derived that are symmetric under particle exchange and involve only first- and second-order correlation functions~\cite{TuraSCIENCE2014}. It was suggested that this could enable the detection of Bell correlations by collective measurements on spin ensembles. Acting on this proposal, we derive a collective witness observable that is tailored to detect Bell correlations in spin-squeezed states of atomic ensembles. We report a measurement of this witness on 480 ultracold Rubidium atoms, revealing Bell correlations in a many-body system.

We derive our Bell correlation witness in the context of a Bell test where $N$ observers $i=1\ldots N$ each repeatedly perform one of two possible local measurements $\mathcal{M}_0^{(i)}$ or $\mathcal{M}_1^{(i)}$ on their part of a composite system, and observe one of two possible outcomes $a_i=\pm1$. For example, the system could be an ensemble of atomic spins where each observer is associated with one atom, and the measurements correspond to spin projections along different axes. When all observers choose to measure $\mathcal{M}_0$,  one determines experimentally the sum of their average outcomes $\mathcal{S}_0=\sum_{i=1}^N \avg{\mathcal{M}_0^{(i)}}$ and correlations $\mathcal{S}_{00}=\sum_{i,j=1 (i\neq j)}^N \avg{\mathcal{M}_0^{(i)}\mathcal{M}_0^{(j)}}$ [see Section~\ref{supp:derivation} of~\cite{supplementary} for a definition in terms of measured frequencies]. Similarly, $\mathcal{S}_{11}=\sum_{i,j=1 (i\neq j)}^N \avg{\mathcal{M}_1^{(i)}\mathcal{M}_1^{(j)}}$ is determined when all observers choose $\mathcal{M}_1$. A more complex correlation $\mathcal{S}_{01}=\sum_{i,j=1 (i\neq j)}^N \avg{\mathcal{M}_0^{(i)}\mathcal{M}_1^{(j)}}$ is quantified by letting all pairs of observers choose opposite measurements, which requires repeated observations of identically prepared states of the system, as some of these measurements are mutually exclusive. In Ref.~\cite{TuraSCIENCE2014} a Bell inequality was derived that contains only these symmetric one- and two-body correlators:
\begin{equation}
	\label{eq:Bellinequality}
	2 \mathcal{S}_0 + \frac{1}{2}\mathcal{S}_{00} + \mathcal{S}_{01} + \frac{1}{2}\mathcal{S}_{11} + 2 N \ge 0.
\end{equation}
If an experiment violates  this inequality, the conditional probabilities $P(a_1,\ldots,a_N|x_1,\ldots,x_N)$ to obtain measurement results $a_1,\ldots,a_N$ for given measurement settings $x_1,\ldots,x_N$ (with $x_i\in\{0,1\}$) cannot be explained by pre-established agreements; \ie\ $P(a_1,\ldots,a_N|x_1,\ldots,x_N) \neq \int\dd\lambda\, P(\lambda)\, P(a_1|x_1,\lambda)\cdots P(a_N|x_N,\lambda)$ where $P(\lambda)$ is the probability of using agreement $\lambda$. In this case, we say that the system is Bell-correlated. For illustration, consider again the situation where each observer performs measurements on the spin of an atom in a large ensemble. If the system is Bell-correlated, appropriate measurements on the atomic spins show statistics that cannot be explained by a recipe that determines the measurement results for each atom independently of the measurement results and settings of the other atoms.

The form of $\mathcal{S}_{01}$ demands that we can set the measurement type of each observer individually. Testing the Bell inequality~\ref{eq:Bellinequality} thus requires more than collective measurements, which are sometimes the only available option in many-body systems. A way around this requirement is to replace the Bell inequality, which guarantees both that the state is Bell correlated and that appropriate measurements were actually performed, by a witness inequality that assumes a quantum-mechanical description and correct experimental calibration of the measurements. A similar approach has been successfully employed to detect entanglement with collective measurements only~\cite{BancalPRL2011,SoerensenNATURE2001,SoerensenPRL2001,GrossNATURE2010,RiedelNATURE2010,HyllusPRA2012}. We associate each observer $i$ with a spin $1/2$ (in our experiment, a pseudo-spin representing two energy levels of an atom). The measurements are spin projections $\mathcal{M}_\vect{d}^{(i)} = 2\vect{\op{s}}^{(i)} \cdot \vect{d}$ along an axis $\vect{d}$, where $2\vect{\op{s}}^{(i)}=\{\op{\sigma}_x^{(i)}, \op{\sigma}_y^{(i)}, \op{\sigma}_z^{(i)}\}$ is the Pauli vector. All other energy levels of the atoms, as well as further degrees of freedom (\eg\ atomic motion), are irrelevant for the measurements. We define the total spin observable $\op{S}_{\vect{d}} = \vect{d}\cdot\sum_{i=1}^N \vect{\op{s}}^{(i)}$ in the direction $\vect{d}$, which can be probed by collective measurements on the entire system. For two unit vectors $\vect{a}$ and $\vect{n}$ we now consider the observable
\begin{equation}
	\label{eq:witness}
	\op{W}=-\left|\frac{\op{S}_{\vect{n}}}{N/2}\right| + (\vect{a}\cdot\vect{n})^2 \frac{\op{S}_{\vect{a}}^2}{N/4} + 1- (\vect{a}\cdot\vect{n})^2,
\end{equation}
defined in terms of total-spin observables only. Setting $\mathcal{M}_\vect{n}^{(i)}=\mathcal{M}_0^{(i)}$ and $\mathcal{M}_\vect{m}^{(i)}=\mathcal{M}_1^{(i)}$ with  $\vect{m}=2(\vect{a}\cdot\vect{n})\vect{a}-\vect{n}$, the expectation value of $\op{W}$ can be reexpressed in terms of one- and two-body correlations functions using $\avg{\op{S}_{\vect{n}}} = \mathcal{S}_0/2$ and $16(\vect{a}\cdot\vect{n})^2\avg{\op{S}_{\vect{a}}^2} = \mathcal{S}_{00} + 2\mathcal{S}_{01} + \mathcal{S}_{11} + 4N(\vect{a}\cdot\vect{n})^2$, see Section~\ref{supp:derivation} of~\cite{supplementary}. The Bell inequality~\ref{eq:Bellinequality} then guarantees that $\avg{\op{W}}\geq 0$ whenever the state of the system is not Bell-correlated. By construction, this Bell correlation witness $\op{W}$ only involves first and second moments of collective spin measurements along two directions $\vect{a}$ and $\vect{n}$, making it well suited for experiments on many-body systems, especially of indistinguishable particles. Although this inequality was derived with assumptions about the measurement settings, it does not make any assumptions about the measured state. In particular, we do not need to assume that the state is symmetric under particle exchange. Moreover, this inequality applies whether the particles  are spatially separated or not, similar to entanglement witnesses~\cite{HyllusPRA2012}, under the common assumption that particles do not communicate (interact) through unknown channels. Although such an assumption would be questioned in a Bell test aimed at disproving the locally causal nature of the world, it is a well-satisfied and common assumption in the present context, where the goal is to explore correlations in a many-body system, assuming quantum mechanics to be valid.

For collective measurements, $N$ is taken to be the number of detected particles, which may fluctuate slightly between experimental runs.   If this is the case, we can replace $N$ in Eq.~\ref{eq:witness} by the observable $\op{N}$, and introduce the scaled collective spin $\mathcal{C}_{\vect{n}}=\avg{2\op{S}_{\vect{n}}/\op{N}}$ and the scaled second moment $\zeta_{\vect{a}}^2 = \avg{4\op{S}_{\vect{a}}^2/\op{N}}$, see Section~\ref{supp:derivation} of~\cite{supplementary}. The inequality then becomes
\begin{equation}
	\label{eq:nonlocalitydirect}
	\mathcal{W} =
    -|\mathcal{C}_{\vect{n}}|
    +(\vect{a}\cdot\vect{n})^2
    \zeta_{\vect{a}}^2
    + 1-(\vect{a}\cdot\vect{n})^2
    \ge 0,
\end{equation}
which is valid for any two axes $\vect{a}$ and $\vect{n}$ and for all non-Bell-correlated states. From this inequality, a criterion follows that will facilitate comparison with well-known spin-squeezing criteria: for any two axes $\vect{a}$ and $\vect{b}$ perpendicular to each other,
\begin{equation}
	\label{eq:nonlocal2}
	\zeta_{\vect{a}}^2 \ge \frac12\left(1-\sqrt{1-\mathcal{C}_{\vect{b}}^2}\right)
\end{equation}
holds for all non-Bell-correlated states [derivation in Section~\ref{supp:perpendicularwitness} of~\cite{supplementary}]. The experiment reported below shows a violation of the inequalities in Eqs.~\ref{eq:nonlocalitydirect} and~\ref{eq:nonlocal2} in an atomic ensemble, hence demonstrating Bell correlations between the atomic spins.

We perform experiments with two-component Bose-Einstein condensates (BECs) of rubidium-87 atoms trapped magnetically on an atom chip~\cite{BoehiNATURE2009} and prepared in a spin-squeezed state as in Refs.~\cite{RiedelNATURE2010,OckeloenPRL2013} [see Section~\ref{supp:experimental} of~\cite{supplementary}]. We start with a BEC without discernible thermal component in the ground-state hyperfine level $\ket{F=1,m_F=-1}\equiv\ket{1}$. We are only concerned with the spin state of the atoms, whereas their uniform motional BEC state is irrelevant for the system's description. We perform only collective manipulations and measurements that are symmetric under particle exchange. A two-photon resonant Rabi field addresses the hyperfine transition from $\ket{1}$ to $\ket{F=2,m_F=1} \equiv \ket{2}$, with these two states representing a pseudo-spin 1/2 for each atom. The internal state of the entire BEC  is described by a collective spin, with the component $\op{S}_z=(\op{N}_1-\op{N}_2)/2$ corresponding to half the atom number difference between the two states. With a $\pi/2$ Rabi pulse we prepare a coherent spin state $[(\ket{1}+\ket{2})/\sqrt{2}]^{\otimes N}$, in which the atomic spins are uncorrelated. To establish correlations between the spins, we make use of elastic collisions, which give rise to a Hamiltonian $\op{H} = \chi \op{S}_z^2$. Controlling the rate $\chi$ with a state-dependent potential~\cite{RiedelNATURE2010}, we evolve the system in time with $\op{H}$ to produce a spin-squeezed state~\cite{KitagawaPRA1993,SoerensenNATURE2001}, which has reduced quantum noise in one collective spin component (Fig.~\ref{fig:data}A). To characterize  this state, we count the numbers of atoms $N_1$ and $N_2$ in the two hyperfine states by resonant absorption imaging~\cite{RiedelNATURE2010}. We correct the data for imaging noise and collisional phase shifts. From averages over many measurements we determine $\mathcal{C}_{\vect{z}}$ and $\zeta_{\vect{z}}^2$. Projections along other spin directions are obtained by  appropriate Rabi rotations before the measurement. 

For the measurement of the Bell correlation witness $\mathcal{W}$ we use BECs with a total atom number of $N=476\pm 21$ (preparation noise after post-selection) and $-5.5(6)\,\mathrm{dB}$ of spin squeezing according to the Wineland criterion~\cite{WinelandPRA1994}. We first measure $\zeta_{\vect{a}}^2$, choosing $\vect{a}$ to be the squeezing axis where $\mathcal{C}_{\vect{a}}\approx 0$ and $\zeta_{\vect{a}}^2$ is minimized (Fig.~\ref{fig:data}B). We find $\zeta_{\vect{a}}^2=0.272(37)$; all quoted uncertainties are statistical standard deviations. For the measurement of $\mathcal{C}_{\vect{n}}$, we sweep the vector $\vect{n}$ in the plane defined by the state's center $\vect{b}$ and the vector $\vect{a}$ (Fig.~\ref{fig:data}A) by applying a Rabi pulse of duration $\tau$. The measurement of $\mathcal{C}_{\vect{n}(\tau)}$ as a function of $\tau$ is shown in Fig.~\ref{fig:data}C. From a sinusoidal fit to the observed Rabi oscillation we obtain the Rabi contrast $\mathcal{C}_{\vect{b}}=0.980(2)$ as well as a precise calibration of $\vect{a}\cdot\vect{n}(\tau)=\cos[\vartheta(\tau)]$  needed to evaluate $\mathcal{W}$. From the resulting measurement of $\mathcal{W}(\tau)$ (Fig.~\ref{fig:data}E), we observe a violation of inequality~\ref{eq:nonlocalitydirect} over a large range of angles. For $\vartheta=128^{\circ}$ we see the strongest violation with a statistical significance of 3.8 standard deviations (red square in Fig.~\ref{fig:data}E).

Inequality~\ref{eq:nonlocalitydirect} relies on a fine balance between competing terms, and a satisfactory demonstration of its violation depends on accurate knowledge of the angle $\vartheta$ between $\vect{a}$ and $\vect{n}$. Inequality~\ref{eq:nonlocal2}, on the other hand, is more robust to uncertainties in this angle, and shows that our entire data set is inconsistent with the hypothesis of our state not being Bell-correlated. The black data point in Fig.~\ref{fig:robustviolation} represents our data set by its Rabi contrast (the amplitude of the red fit in Fig.~\ref{fig:data}C) and squeezed second moment (Fig.~\ref{fig:data}B), giving an overall likelihood of 99.9\% for Bell correlations [see Section~\ref{supp:experimental} of~\cite{supplementary}]. This likelihood can be interpreted as a $p$-value of 0.1\% for excluding the hypothesis: ``Our data were generated by a state that has no Bell correlations, in the presence of Gaussian noise.'' An experiment closing the statistics loophole would exclude all possible non-Bell-correlated states, including those producing statistics with rare events. However, because of the way the bounds on $\mathcal{W}$ vary with $N$, such an experiment would require a number of measurements that increases with the number of spins [see Section~\ref{supp:statistics} of~\cite{supplementary}].

We now discuss how our Bell correlation witness is connected to entanglement measures that were used previously to characterize spin-squeezed BECs~\cite{SoerensenNATURE2001,SoerensenPRL2001,GrossNATURE2010,RiedelNATURE2010,HyllusPRA2012}. These entanglement measures depend on the squeezed variance, for which the squeezed second moment $\zeta_{\vect{a}}^2$ is an upper bound (with equality if $\mathcal{C}_{\vect{a}}=0$, which is close to what we have measured). In terms of the latter, the Wineland spin-squeezing parameter~\cite{WinelandPRA1994} $\xi^2\le\zeta_{\vect{a}}^2/\mathcal{C}_{\vect{b}}^2$ witnesses entanglement~\cite{SoerensenNATURE2001} if $\xi^2<1$, shown as a red shaded region in Fig.~\ref{fig:robustviolation}. Similarly, $(k\!+\!1)$-particle entanglement is witnessed by measuring squeezed variances (and hence $\zeta_{\vect{a}}^2$) below the gray $k$-producibility curves~\cite{SoerensenPRL2001} in Fig.~\ref{fig:robustviolation}. We note that these entanglement witnesses refer to the Ramsey contrast, whereas our data point in Fig.~\ref{fig:robustviolation} refers to the measured Rabi contrast; in our experiment these two quantities have nearly identical values. We can thus draw conclusions about both entanglement and Bell correlations from Fig.~\ref{fig:robustviolation}. In particular, we conclude that our witness requires at least 3\,dB of spin squeezing for detecting Bell correlations.

We have shown that Bell correlations can be created and detected in many-body systems. This result has been obtained from a witness that requires collective measurements only. Although we have tested this witness with a spin-squeezed BEC, it could also be tested on other systems such as thermal atoms in a spin-squeezed state. Our results imply that the correlations between the atoms in a spin-squeezed Bose-Einstein condensate are strong enough to violate a Bell inequality. This Bell inequality could be violated directly by first localizing the atoms, \eg\ through a non-destructive, spin-independent measurement of their position, and then measuring their internal states individually [see Section~\ref{supp:BellTest} of~\cite{supplementary}]. Further study of these states may enable insights into many-body correlations outside of the quantum formalism. Our results naturally raise the question of how our witness can be extended to detect genuine multipartite nonlocality~\cite{SvetlichnyPRD1987} or to quantify the degree of nonlocality~\cite{BancalPRL2009,CurchodPRA2015}, in a similar way as the degree of entanglement can be quantified in terms of $k$-producibility (Fig.~\ref{fig:robustviolation})~\cite{SoerensenPRL2001}. Finally, Bell correlations are a resource in quantum information theory, \eg\ for certifiable randomness generation. Although Bell-correlation-based randomness has been extracted from two-qubit systems~\cite{PironioNATURE2010}, an implementation in a many-body system would considerably increase the amount of randomness per experimental run.

\section*{Acknowledgments}

We thank Remik Augusiak, Peter Drummond and Maciej Lewenstein for fruitful discussions. RS, JDB, BA, MF, PT and NS acknowledge support from the Swiss National Science Foundation (SNSF) through grants PP00P2-150579, 20020-149901 and NCCR QSIT. PT acknowledges support from the European Union through project SIQS. NS acknowledges the Army Research Laboratory Center for Distributed Quantum Information via the project SciNet. JDB and VS acknowledge funding from the Singapore Ministry of Education (partly through Academic Research Fund Tier 3 MOE2012-T3-1-009) and the National Research Foundation of Singapore.

\section*{Author Contributions}

NS and JDB initiated the study, with input from RS. JDB derived optimal Bell inequalities and performed statistics tests discussing with RS, NS and VS. RS and PT derived the witnesses from the Bell inequality, with support from JDB and NS. BA, MF and RS performed experiments and analyzed data, supervised by PT. All authors discussed the results and contributed to the manuscript.

\clearpage
\begin{figure}[!ht]
	\begin{center}
		\includegraphics[width=0.9\textwidth]{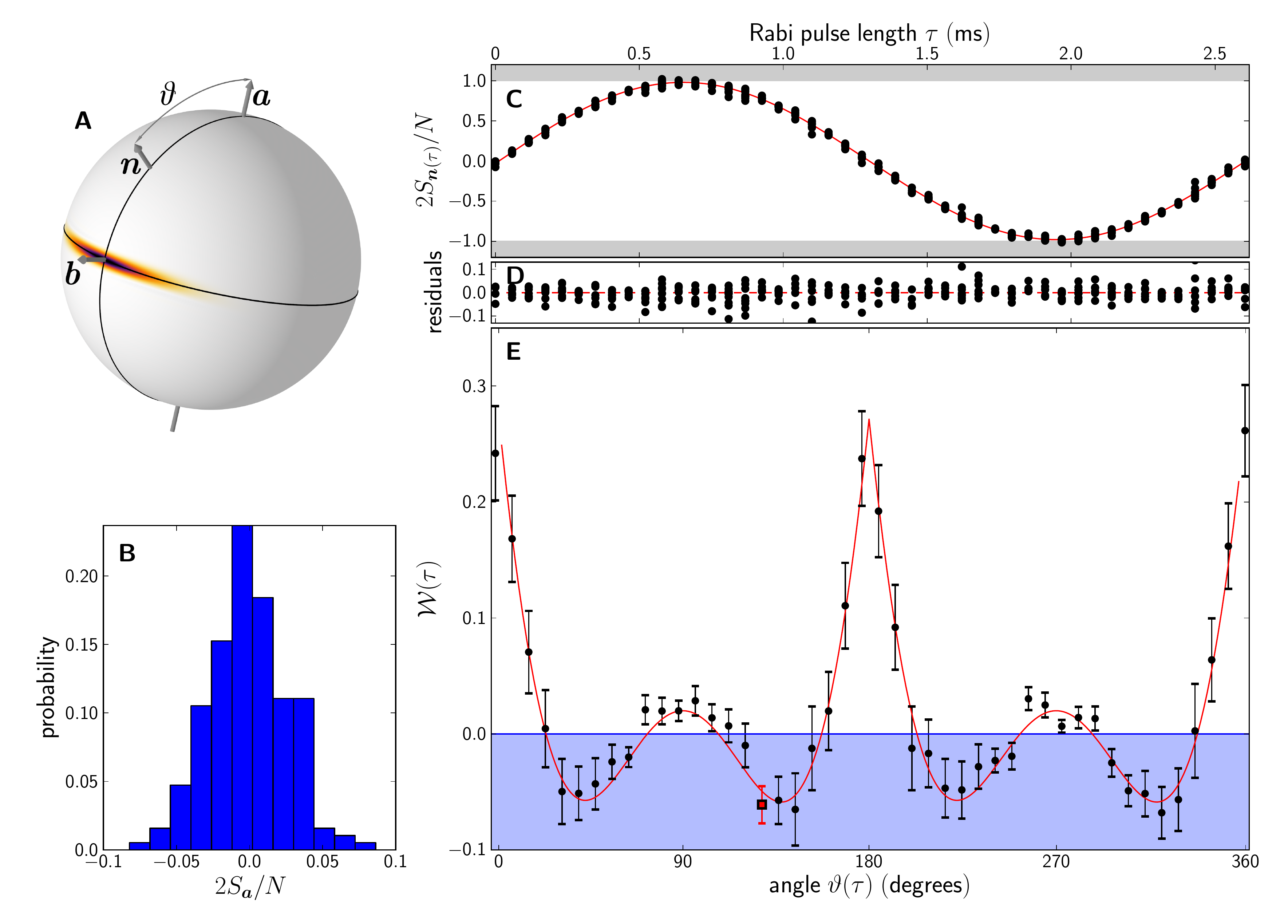}
		\caption{\label{fig:data}
		\textbf{Observation of Bell correlations in a BEC with inequality~\ref{eq:nonlocalitydirect}.}
		\textbf{(A)} Illustration of the spin-squeezed state [Wigner function~\cite{SchmiedNJP2011}] and the axes used in the measurement of the Bell correlation witness $\mathcal{W}$.  The vector $\vect{n}$ lies in the plane spanned by the squeezing  axis $\vect{a}$  and the state's center $\vect{b}$. The squeezing and anti-squeezing planes are indicated with thin black lines.
		\textbf{(B)} Histogram of measurements of $2S_{\vect{a}}/N$ from which we determine $\zeta_{\vect{a}}^2$.
		\textbf{(C)} Individual measurements of $2S_{\vect{n}(\tau)}/N$ as a function of Rabi pulse length $\tau$. The red line is a sinusoidal fit from which we determine the Rabi contrast and $\vect{a}\cdot\vect{n}(\tau)=\cos[\vartheta(\tau)]$, see Section~\ref{supp:experimental} of~\cite{supplementary}.
		\textbf{(D)} Residuals of the fit of (C).
		\textbf{(E)} Measurement of $\mathcal{W}(\tau)$ as a function of $\vartheta(\tau)$. The red continuous line is the value of $\mathcal{W}(\tau)$ computed from the measurement of $\zeta_{\vect{a}}^2$ and the fitted Rabi oscillation [red line in (C)]. Bell correlations are present in the blue shaded region. Note that the observed four-fold symmetry of $\mathcal{W}(\tau)$ indicates that $\vect{a}\cdot\vect{n}(\tau)$ is well calibrated. The red square data point at $\vartheta=128^{\circ}$ violates inequality~\ref{eq:nonlocalitydirect} by 3.8 standard deviations.}
	\end{center}
\end{figure}

\clearpage
\begin{figure}[!ht]
	\begin{center}
		\includegraphics[width=0.6\textwidth]{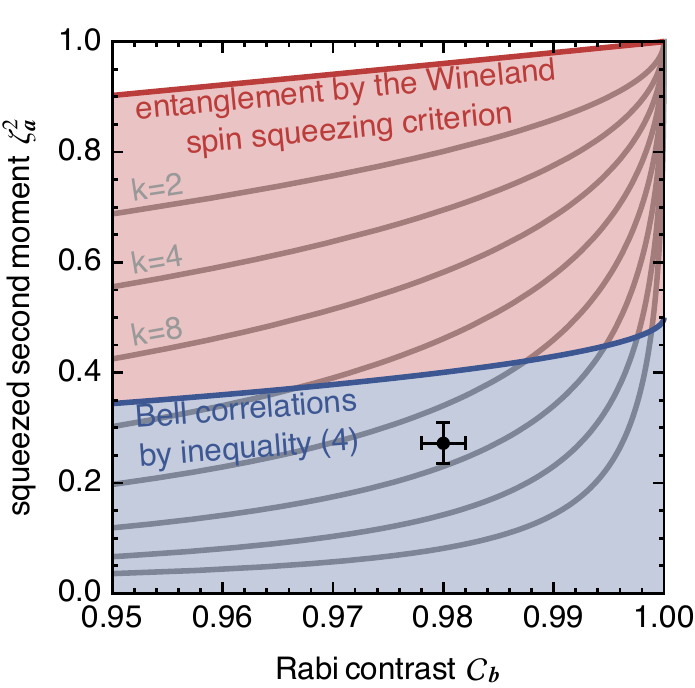}
		\caption{\label{fig:robustviolation}
		\textbf{Observation of Bell correlations in a BEC with inequality~\ref{eq:nonlocal2}.}
		Black circle: the data set of Fig.~\ref{fig:data} expressed in terms of the Rabi contrast $\mathcal{C}_{\vect{b}}$ and the squeezed second moment $\zeta_{\vect{a}}^2$, with $1\sigma$ error bars. 
		Blue shaded region: Bell correlations detected by violation of inequality~\ref{eq:nonlocal2}. A pair of random variables with the same parameters as our data set has a 99.9\% overlap with this region [see Section~\ref{supp:experimental} of~\cite{supplementary}].
		Red shaded region: entanglement witnessed by spin squeezing~\cite{WinelandPRA1994,SoerensenNATURE2001}.
		Gray lines: limits on $\zeta_{\vect{a}}^2$ below which there is at least $(k\!+\!1)$-particle entanglement~\cite{SoerensenPRL2001}, increasing in powers of two up to $k=256$.
		Our data set has a 99\% overlap with the area below the limit for $k=24$-particle entanglement.}
	\end{center}
\end{figure}


\clearpage
\setcounter{figure}{0}
\renewcommand\thefigure{S\arabic{figure}}
\setcounter{equation}{0}
\renewcommand\theequation{S\arabic{equation}}
\section*{Bell Correlations in a Bose-Einstein Condensate\\Supplementary Materials}

\section{Detailed derivation of the Bell correlation witness}
\label{supp:derivation}

In general, Bell inequalities are tested in scenarios where a number $N$ of observers each perform one of several possible measurements and observe discrete outcomes. We label the measurement setting of party $i\in\{1,\ldots,N\}$ as $x_i$ and its outcome as $a_i=\pm1$. For a given set of measurement settings $\bar{x}=(x_1,\ldots,x_N)$ the experiment is repeated many times to measure the frequencies (conditional probabilities) $P(\bar{a}|\bar{x})$ of observed outcomes $\bar{a}=(a_1,\ldots,a_N)$. A Bell inequality is then a linear combination of such conditional probabilities, together with a bound satisfied by every classical model (relying on pre-established agreements)~\cite{BrunnerRMP2014}.

We define the marginal probability distribution for subsystem $i$, $P_i(a_i|x_i)=\sum_{\bar{a}'|a'_i=a_i}P(\bar{a}'|\bar{x})$, and similarly the bipartite marginal for subsystems $i$ and $j$, $P_{ij}(a_i,a_j|x_i,x_j)=\sum_{\bar{a}'|a'_i=a_i \land a'_j=a_j}P(\bar{a}'|\bar{x})$. We notice that these marginals do not depend on the other elements of $\bar{x}$ because of the no-signaling principle. In terms of these marginals, the expectation values in inequality~\ref{eq:Bellinequality} are $\mathcal{S}_0 = \sum_{i=1}^N \sum_{a=\pm1} a\, P_i(a|0)$ and $\mathcal{S}_{k\ell} = \sum_{i,j=1 (i\neq j)}^N \sum_{a,b=\pm1} a b\, P_{ij}(a,b|k,\ell)$.

The expectation values of the spin operators in Eq.~\ref{eq:witness} are given in the last paragraph of Ref.~\cite{TuraSCIENCE2014}. With $\vect{m}=2(\vect{a}\cdot\vect{n})\vect{a}-\vect{n}$ such that $\|\vect{a}\|=\|\vect{m}\|=\|\vect{n}\|=1$, they become
\begin{eqnarray}
	\mathcal{S}_0 &=& 2\avg{\op{S}_{\vect{n}}} \nonumber\\
	\mathcal{S}_{00} &=& 4\avg{\op{S}_{\vect{n}}^2}-N \nonumber\\
	\mathcal{S}_{11} &=& 4\avg{\op{S}_{\vect{m}}^2}-N \nonumber\\
		&=& 16(\vect{a}\cdot\vect{n})^2\avg{\op{S}_{\vect{a}}^2}-8(\vect{a}\cdot\vect{n})\avg{\op{S}_{\vect{a}}\op{S}_{\vect{n}}+\op{S}_{\vect{n}}\op{S}_{\vect{a}}}+4\avg{\op{S}_{\vect{n}}^2}-N \nonumber\\
	\mathcal{S}_{01} &=& \avg{(\op{S}_{\vect{n}}+\op{S}_{\vect{m}})^2}-\avg{(\op{S}_{\vect{n}}-\op{S}_{\vect{m}})^2}-N(\vect{n}\cdot\vect{m})\nonumber\\
		&=& 4(\vect{a}\cdot\vect{n})\avg{\op{S}_{\vect{a}}\op{S}_{\vect{n}}+\op{S}_{\vect{n}}\op{S}_{\vect{a}}}-4\avg{\op{S}_{\vect{n}}^2}
			- N\left[2(\vect{a}\cdot\vect{n})^2-1\right],
\end{eqnarray}
and thus $\mathcal{S}_{00}+2\mathcal{S}_{01}+\mathcal{S}_{11}=16(\vect{a}\cdot\vect{n})^2\avg{\op{S}_{\vect{a}}^2}-4N(\vect{a}\cdot\vect{n})^2$ is independent of $\op{S}_{\vect{n}}$. With these relations and the fact that $\op{S}_{-\vect{n}}=-\op{S}_{\vect{n}}$ we find that inequality~\ref{eq:Bellinequality} proves that
\begin{equation}
	\label{eq:witnessN}
	-\left|\frac{\avg{\op{S}_{\vect{n}}}}{N/2}\right| + (\vect{a}\cdot\vect{n})^2\frac{\avg{\op{S}_{\vect{a}}^2}}{N/4}+1-(\vect{a}\cdot\vect{n})^2 \ge 0
\end{equation}
for every state that is not Bell-correlated.

If the total atom number $N$ varies from one measurement to the next, we extend the above analysis by replacing $\avg{\op{S}_{\vect{n}}}/N$ by $\avg{\op{S}_{\vect{n}}/\op{N}}$ and $\avg{\op{S}_{\vect{a}}^2}/N$ by $\avg{\op{S}_{\vect{a}}^2/\op{N}}$, which is possible because the spin operators $\op{S}_{\vect{n}}$ and $\op{S}_{\vect{a}}$ commute with the atom number operator $\op{N}$. More precisely, note that the operators $\op{S}_{\vect{n}}$ and $\op{S}_{\vect{a}}$ are block-diagonal in terms of the particle number, \ie\ $\op{S}_{\vect{n}}=\underset{N\geq 1}{\oplus} \op{S}_{\vect{n}}^{(N)}$ where $\op{S}_{\vect{n}}^{(N)}$ denotes the action of the spin operator $\op{S}_{\vect{n}}$ on the subspace with particle number $N$, and similarly for $\op{S}_{\vect{a}}$. In this basis, the particle number operator takes the form $\op{N}=\underset{N\geq 1}{\oplus} N\cdot \one^{(N)}$, where $\one^{(N)}$ is the identity operator on the space with particle number $N$, and $\one=\underset{N\geq 1}{\oplus} \one^{(N)}$. Therefore, we have that $\avg{\op{S}_{\vect{n}}/\op{N}} \equiv \avg{(\underset{N\geq 1}{\oplus}\op{S}_{\vect{n}}^{(N)})\cdot (\underset{N\geq 1}{\oplus} N\cdot \one^{(N)})^{-1}}=\sum_N p_N \avg{\op{S}_{\vect{n}}}_N/N$, where $\avg{\op{S}_{\vect{n}}}_N=\mathrm{Tr}(\op{S}_{\vect{n}} \op{\rho}^{(N)})$ is the $N$-particle contribution to the spin operator, $\op{\rho}^{(N)}=(\one^{(N)}\op{\rho}\,\one^{(N)})/p_N$ is the normalized $N$-particle component of the density matrix $\op{\rho}$, and $p_N=\mathrm{Tr}(\one^{(N)}\op{\rho}\,\one^{(N)})$ is the probability of having $N$ particles. Similarly, we can write $\avg{\op{S}_{\vect{a}}^2/\op{N}}=\sum_N p_N \avg{\op{S}_{\vect{a}}^2}_N/N$ with $\avg{\op{S}_{\vect{a}}^2}_N=\mathrm{Tr}(\op{S}_{\vect{a}}^2\op{\rho}^{(N)})$, so that
\begin{equation}
	-\left|\left<\frac{\op{S}_{\vect{n}}}{\op{N}/2}\right>\right|
	+ (\vect{a}\cdot\vect{n})^2 \left<\frac{\op{S}_{\vect{a}}^2}{\op{N}/4}\right>
	+1-(\vect{a}\cdot\vect{n})^2
	= \sum_N p_N \left[
	-\left|\frac{\avg{\op{S}_{\vect{n}}}_N}{N/2}\right|
	+ (\vect{a}\cdot\vect{n})^2 \frac{\avg{\op{S}_{\vect{a}}^2}_N}{N/4}
	+1-(\vect{a}\cdot\vect{n})^2
	\right].
\end{equation}
If no component $\op{\rho}^{(N)}$ is Bell-correlated, then every term in this sum is nonnegative according to inequality~\ref{eq:witnessN}, and consequently inequality~\ref{eq:nonlocalitydirect} is satisfied. Conversely, a violation of inequality~\ref{eq:nonlocalitydirect} proves that at least one component $\op{\rho}^{(N)}$ of the system's state is Bell-correlated.

\subsection{A Bell correlation witness for two perpendicular measurement directions}
\label{supp:perpendicularwitness}

Inequality~\ref{eq:nonlocalitydirect} relies on a fine balance between competing terms, and a satisfactory demonstration of its violation depends on accurate knowledge of the angle $\vartheta$ between $\vect{a}$ and $\vect{n}$ (see Section~\ref{supp:experimental} for calibration procedure). Here we derive a Bell correlation witness that is more robust to uncertainties in this angle, that summarizes the overall violation in a single comparison (see Fig.~\ref{fig:robustviolation}), and that can be compared to some known entanglement witnesses.

We decompose $\vect{n}=\vect{a}\cos(\vartheta)+\vect{b}\sin(\vartheta)\cos(\varphi)+\vect{c}\sin(\vartheta)\sin(\varphi)$ in terms of three ortho-normal vectors $\{\vect{a},\vect{b},\vect{c}\}$ (see Fig.~\ref{fig:data}A for an example, with $\vect{c}=\vect{a}\times\vect{b}$), and define the scaled collective spin components $\mathcal{C}_{\vect{a}}=\avg{2\op{S}_{\vect{a}}/\op{N}}$ etc. The resulting inequality $\zeta_{\vect{a}}^2 \ge -[\mathcal{C}_{\vect{a}}\cos(\vartheta)+\mathcal{C}_{\vect{b}}\sin(\vartheta)\cos(\varphi)+\mathcal{C}_{\vect{c}}\sin(\vartheta)\sin(\varphi)+\sin^2(\vartheta)]/\cos^2(\vartheta)$, valid for all non-Bell-correlated states according to inequality~\ref{eq:nonlocalitydirect}, can be violated if there exists an angular direction $(\vartheta,\varphi)$ for which it is violated; that is, the measurements along the perpendicular axes $\{\vect{a},\vect{b},\vect{c}\}$ on any non-Bell-correlated state satisfy
\begin{eqnarray}
	\label{eq:nonlocab}
	\zeta_{\vect{a}}^2 &\ge& Z(\mathcal{C}_{\vect{b}\vect{c}},\mathcal{C}_{\vect{a}})
	= \max_{\vartheta\in[0,\pi]} \left[\frac{\mathcal{C}_{\vect{b}\vect{c}}\sin(\vartheta)-\mathcal{C}_{\vect{a}}\cos(\vartheta)-\sin^2(\vartheta)}{\cos^2(\vartheta)}\right]\nonumber\\
	&\ge& Z(\mathcal{C}_{\vect{b}},0) = \frac{1-\sqrt{1-\mathcal{C}_{\vect{b}}^2}}{2}
\end{eqnarray}
with $\mathcal{C}_{\vect{b}\vect{c}}=\sqrt{\mathcal{C}_{\vect{b}}^2+\mathcal{C}_{\vect{c}}^2}$. The function $Z(\mathcal{C}_{\vect{b}\vect{c}},\mathcal{C}_{\vect{a}})$ is discussed in more detail below. All non-Bell-correlated states thus satisfy inequality~\ref{eq:nonlocal2}, which we have violated experimentally as shown in Fig.~\ref{fig:robustviolation}.

\subsection{Details on the function $Z(\mathcal{C}_{\vect{b}\vect{c}},\mathcal{C}_{\vect{a}})$}

\begin{figure}
	\begin{center}
		\includegraphics[width=0.6\textwidth]{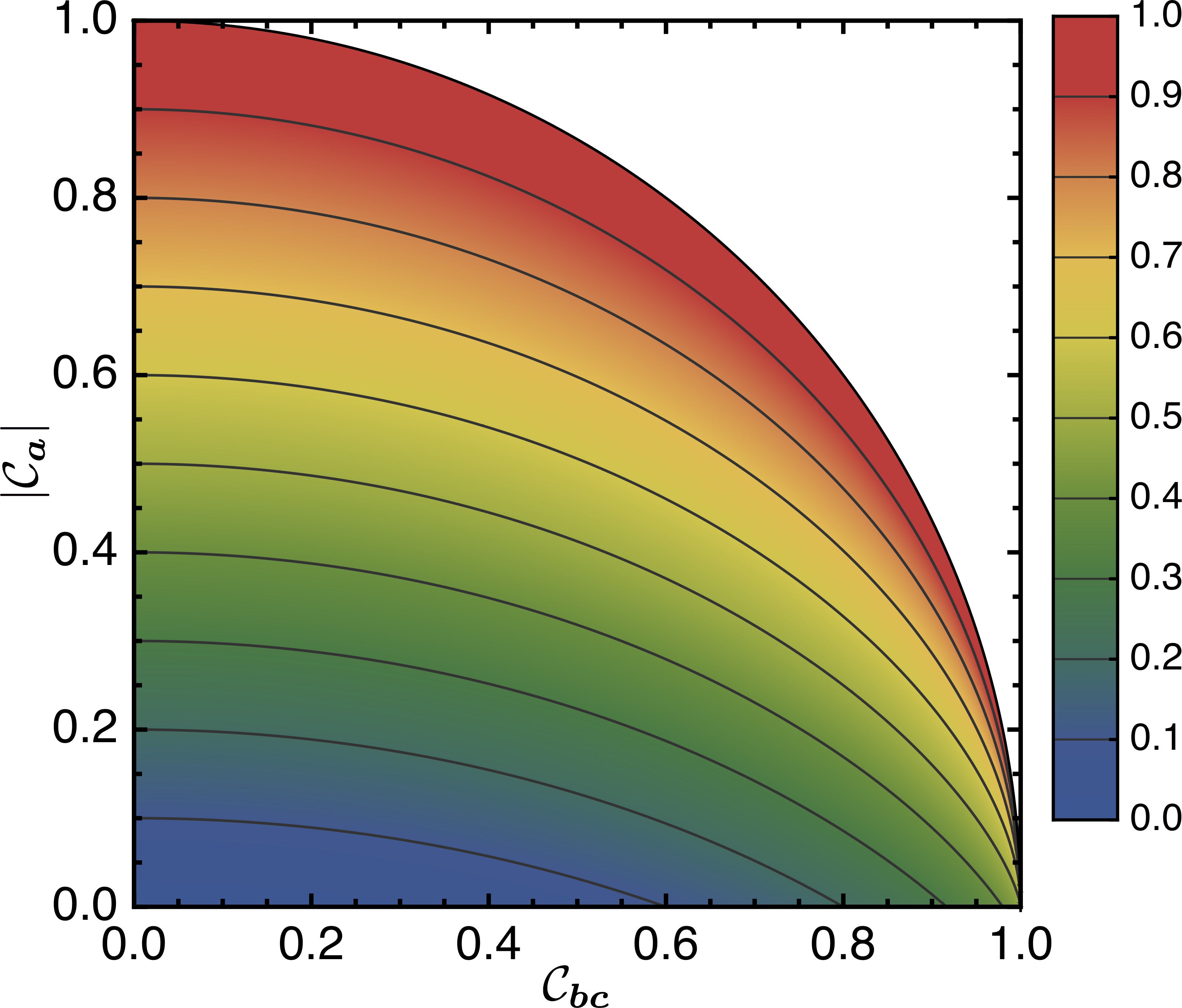}
		\caption{\label{fig:CaCblimit}
		\textbf{Graphical representation of the function $Z(\mathcal{C}_{\vect{b}\vect{c}},\mathcal{C}_{\vect{a}})$ from Eq.~\ref{eq:nonlocab}}.
		The horizontal line $\mathcal{C}_{\vect{a}}=0$ is shown as a blue line in Fig.~\ref{fig:robustviolation}.}
	\end{center}
\end{figure}

For $\mathcal{C}_{\vect{a}}^2+\mathcal{C}_{\vect{b}\vect{c}}^2\le1$, $Z(\mathcal{C}_{\vect{b}\vect{c}},\mathcal{C}_{\vect{a}})$ of Eq.~\ref{eq:nonlocab} satisfies
\begin{eqnarray}
	\label{eq:zetaroot}
	\mathcal{C}_{\vect{b}\vect{c}}^6 + [\mathcal{C}_{\vect{a}}^2 + 4 (1 - Z)]^2 (\mathcal{C}_{\vect{a}}^2 - Z^2) + \mathcal{C}_{\vect{b}\vect{c}}^4 [3 \mathcal{C}_{\vect{a}}^2 + 8 Z (Z - 1) - 1]&\nonumber\\
	+ \mathcal{C}_{\vect{b}\vect{c}}^2 [3 \mathcal{C}_{\vect{a}}^4 - 2 \mathcal{C}_{\vect{a}}^2 (10 Z^2 - 19 Z + 10) + 8 Z (Z - 1) (2 Z^2 - 2 Z - 1)]&=&0,
\end{eqnarray}
and we can find $Z$ numerically as the larger of the two real roots of this polynomial. While explicit formulas exist for this root with validities in different domains of $(\mathcal{C}_{\vect{b}\vect{c}},\mathcal{C}_{\vect{a}})$, they are very long and not suited for printing here. In Fig.~\ref{fig:CaCblimit} we show $Z(\mathcal{C}_{\vect{b}\vect{c}},\mathcal{C}_{\vect{a}})$ graphically. We note that $\partial Z(\mathcal{C}_{\vect{b}\vect{c}},\mathcal{C}_{\vect{a}})/\partial|\mathcal{C}_{\vect{b}\vect{c}}|\ge0$ and $\partial Z(\mathcal{C}_{\vect{b}\vect{c}},\mathcal{C}_{\vect{a}})/\partial|\mathcal{C}_{\vect{a}}|\ge1$, which yields the monotonicity used in inequality~\ref{eq:nonlocab}.

\section{Details on the experimental setup and procedure}
\label{supp:experimental}

We experiment with two-component Bose-Einstein condensates of $^{87}$Rb atoms~\cite{BoehiNATURE2009,RiedelNATURE2010,GrossNATURE2010,OckeloenPRL2013} in the hyperfine states $\ket{F=1,m_F=-1}\equiv\ket{1}$ and $\ket{F=2, m_F=1}\equiv\ket{2}$. The atoms are magnetically trapped on an atom chip~\cite{BoehiNATURE2009} with trapping frequencies $f_x=110$\,Hz and $f_y=f_z=729$\,Hz. The experiment prepares condensates without discernible thermal components and with $N=474\pm27$ atoms (rms preparation noise) in state $\ket{1}$. To remove outliers, we post-select BECs with $N\in[425,520]$. The resulting data set has $N=476\pm21$. Because the magnetic moments of states $\ket{1}$ and $\ket{2}$ are almost equal, the two states experience nearly the same magnetic trapping potential and show very good coherence properties~\cite{BoehiNATURE2009}. The internal state of the atoms is manipulated via two-photon [radio-frequency (rf) and microwave (mw)] resonant Rabi rotations with a Rabi frequency of 384\,Hz, for which the mw field is blue-detuned by $\approx$500\,kHz from the intermediate state $\ket{F=2, m_F=0}$~\cite{OckeloenPRL2013}. Such Rabi pulses achieve coherent population transfers with very high fidelities; in particular, we are not able to detect any atoms transferred to other states of the ground-state manifold. This allows us to focus entirely on the effective two-level system of states $\ket{1}$ and $\ket{2}$, described by pseudo-spins 1/2.

\begin{figure}
	\begin{center}
		\includegraphics[width=0.9\textwidth]{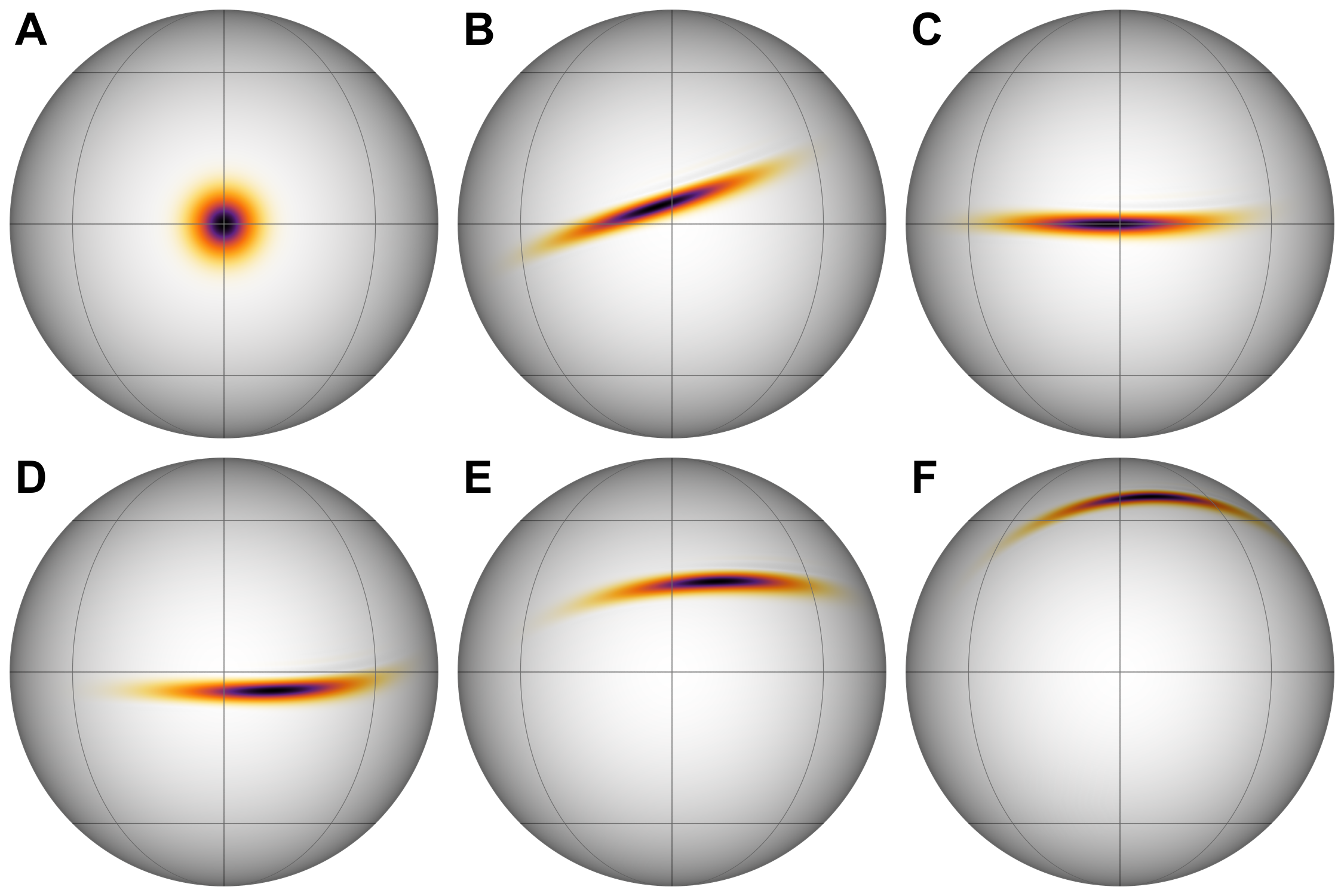}
		\caption{\label{fig:stateprep}
		\textbf{Graphical representation of the experimental sequence for detecting Bell correlations.}
		Spherical projections of the simulated Wigner function on the Bloch sphere~\cite{SchmiedNJP2011} during state preparation and measurement, for a total spin $S=50$ corresponding to $N=2S=100$ pseudo-spin-1/2 particles. Projective measurements are along the vertical ($+z$) spin axis.
		\textbf{(A)} Initial coherent state along the $+x$ spin direction.
		\textbf{(B)} After one-axis twisting, including phase error (left/right) and amplitude error (up/down).
		\textbf{(C)} Measurement of $\zeta^{2}_{\vect{a}}$ after rotation $\mathcal{R}(\kappa,\varphi_0)$ such that $\mathcal{C}_{\vect{a}}\approx0$ and $\zeta^{2}_{\vect{a}}$ is minimal.
		\textbf{(D-F)} Measurements of $\mathcal{C}_{\vect{n}(\tau)}$ for three different Rabi pulse durations $\tau$, corresponding to three different axes $\vect{n}(\tau)$. The difference between (C) and (D) is that the latter contains an additional phase error in the state preparation, which we account for in the calibration of $\vect{a}\cdot\vect{n}(\tau)=\cos[\vartheta(\tau)]$.}
        \end{center}
\end{figure}

The experimental sequence is illustrated in Fig.~\ref{fig:stateprep}, representing the Wigner function of the collective spin state on a sphere~\cite{SchmiedNJP2011}. It starts with the preparation of a spin-squeezed state using the one-axis twisting scheme~\cite{KitagawaPRA1993}. In this scheme, a coherent spin state (Fig.~\ref{fig:stateprep}A) evolves in time with a Hamiltonian $\op{H} = \chi \op{S}_z^2$, resulting in a spin-squeezed state with reduced quantum noise in a certain spin component (Fig.~\ref{fig:stateprep}B). To generate this Hamiltonian, we control the collisional interactions between the two states with a state-dependent microwave near-field potential~\cite{BoehiNATURE2009}. A state-selective splitting of the two potential minima by 150\,nm induces coherent demixing-remixing dynamics as described in Refs.~\cite{RiedelNATURE2010,OckeloenPRL2013}. The preparation needs two complete oscillations ($\approx$55\,ms) to generate --5.5(6)\,dB of spin squeezing according to the Wineland criterion~\cite{WinelandPRA1994}.

The optimization and calibration processes for our state-selective absorption imaging are explained in Refs.~\cite{OckeloenPRL2013,OckeloenPhD2014}. The times of flight are 4.0\,ms for state $\ket{2}$ and 5.5\,ms for $\ket{1}$, and we achieve detection noise levels of $\sigma_{N_1,\mathrm{det}}=4.5$ and $\sigma_{N_2,\mathrm{det}}=3.9$ atoms, where $\sigma_{N_i,\mathrm{det}}^{2}$ are the variances of the measured atom numbers $N_{i,\mathrm{det}}$ due to imaging noise. Starting from the method described in Refs.~\cite{OckeloenPRL2013,OckeloenPhD2014} we add an empirical correction to account for the relatively high optical density of the cloud (which is of order 1). By driving Rabi rotations on BECs with different atom numbers we fit quadratic corrections $\nu_i$ that restore the sinusoidal shapes of the Rabi oscillation for both states. The atom numbers obtained in this way are $N_i=N_{i,\mathrm{det}} + \nu_i N_{i,\mathrm{det}}^{2}$ with $\nu_1=1.46(9)\times10^{-4}$ and $\nu_2=2.57(9)\times10^{-4}$. Omitting these corrections would underestimate the atom number noise. The accuracy of the resulting atom number calibration is confirmed by recording atomic projection noise as a function of atom number~\cite{OckeloenPRL2013,OckeloenPhD2014}, which shows the expected value of $\zeta_{\vect{a}}^2\approx 1$ for coherent states (Fig.~\ref{fig:stateprep}A) independently of $N$.

As illustrated schematically in Fig.~\ref{fig:stateprep}B, our spin-squeezed states are (i) not exactly on the equator, due to different particle loss rates in the two states during the preparation, and (ii) tilted by $\approx$11.0$^{\circ}$ against the horizontal due to the one-axis twisting dynamics. All subsequent operations on these states are resonant Rabi rotations $\mathcal{R}(\alpha,\varphi)$ corresponding to active rotations of the state on the Bloch sphere by an angle $\alpha$ around an axis $\{\cos(\varphi),\sin(\varphi),0\}$ lying in the equatorial spin plane. Projective measurements of the collective pseudo-spin state are always taken along the $+z$ spin axis, but after coherent Rabi rotations applied to the state; in this way, the passive rotations of inequalities~\ref{eq:nonlocalitydirect} and~\ref{eq:nonlocal2} (\ie\ measurements along different axes for a fixed state) are experimentally replaced by active rotations (\ie\ measurements along a fixed axis for differently rotated states).

Inequalities~\ref{eq:nonlocalitydirect} and~\ref{eq:nonlocal2} are most easily violated when $\vect{a}$ is the axis that minimizes $\zeta^{2}_{\vect{a}}$, which in our case means that the measurements should be taken along the squeezed spin component and that the state should be placed on the equator of the sphere as precisely as possible, as illustrated in Fig.~\ref{fig:stateprep}C. In a first set of measurements, we thus rotate the squeezed state by $\mathcal{R}(\kappa,\varphi_0)$, adjusting the pulse area $\kappa$ and the phase $\varphi_0$ so that $\mathcal{C}_{\vect{a}}$ is as close as possible to zero and $\zeta^{2}_{\vect{a}}$ is minimal. In practice,  we scan $\varphi_0$, keeping $\kappa = 11.0^{\circ}$ fixed, and pick the value of $\varphi_0$ that minimizes $|\mathcal{C}_{\vect{a}}|$.

\begin{figure}
	\begin{center}
		\includegraphics[width=0.9\textwidth]{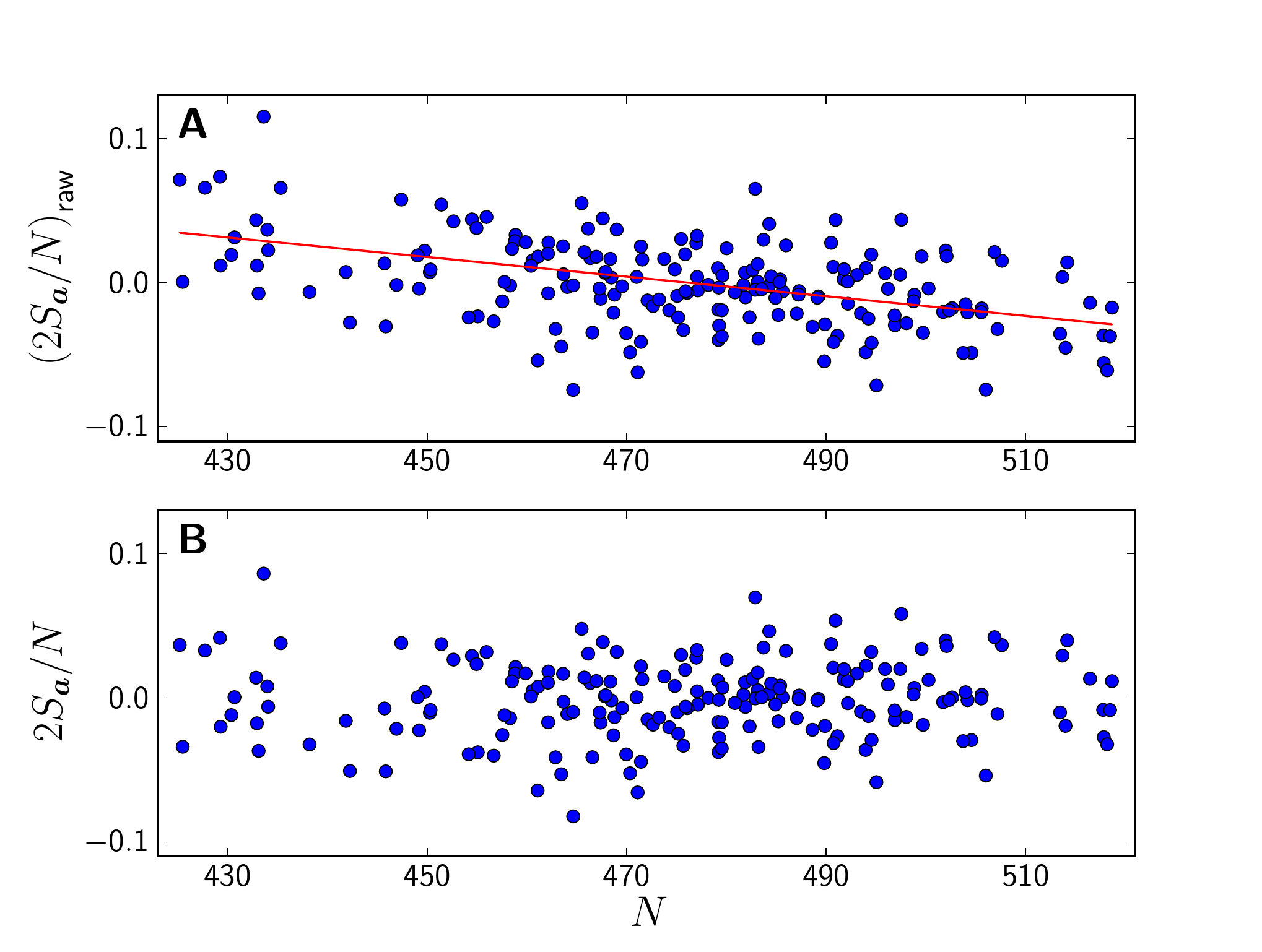}
		\caption{\label{fig:CScorrection}
		\textbf{Experimental data used to estimate $\zeta_{\vect{a}}^2$.}
		\textbf{(A)} Raw measurements $(2S_{\vect{a}}/N)\si{raw}=\frac{N_1-N_2}{N_1+N_2}$, post-selected to $425\le N_1+N_2\le520$.
		The linear trend due to collisional phase shifts (red line), with slope $\lambda=-6.8(12)\times10^{-4}$ per atom determined from a much larger data set, is used to correct the measurements without changing their mean value.
		\textbf{(B)} Corrected measurements $2S_{\vect{a}}/N=(2S_{\vect{a}}/N)\si{raw}-\lambda(N-\avg{N})$.}
	\end{center}
\end{figure}

During the preparation of the spin-squeezed state, collisions lead to an $N$-dependent phase shift between $\ket{1}$ and $\ket{2}$, also know as ``clock shift'' in the context of precision measurements~\cite{OckeloenPhD2014}. Because $N$ fluctuates from shot to shot, this leads to a corresponding fluctuation in the azimuthal position of the state on the Bloch sphere. After applying the rotation $\mathcal{R}(\kappa,\varphi_0)$, this translates into a weak dependence of the measured raw values $(2S_{\vect{a}}/N)\si{raw}=\frac{N_1-N_2}{N_1+N_2}$ on the measured total atom numbers $N=N_1+N_2$, see Fig.~\ref{fig:CScorrection}A. Since this correlation is deterministic and $N$ is measured accurately in every shot, we can correct for this slight misalignment of the measurement axes, similar to what is commonly done in precision atomic clocks. Using a much larger data set we estimate the slope of this correlation to be $\lambda=-6.8(12)\times10^{-4}$ per atom. This value is then used to correct each individual measurement to $2S_{\vect{a}}/N=(2S_{\vect{a}}/N)\si{raw}-\lambda(N-\avg{N})$ without changing their mean values. The corrected measurements $2S_{\vect{a}}/N$ are presented in Fig.~\ref{fig:CScorrection}B. A histogram of the same data is shown in Fig.~\ref{fig:data}B. For a precise determination of $\zeta_{\vect{a}}^2$ we also correct our data for detection noise: the best estimate is obtained by subtracting $(\sigma_{N_1,\mathrm{det}}^{2}+\sigma_{N_2,\mathrm{det}}^{2})/\avg{N}$ from $\avg{(2S_{\vect{a}}/N)^2\times N}$. We obtain $\mathcal{C}_{\vect{a}}=\avg{2S_{\vect{a}}/N}=-1(2)\times10^{-3}$ and $\zeta_{\vect{a}}^{2}=0.272(37)$, with a sample size of 190 points. If we do not subtract detection noise for the estimate of $\zeta_{\vect{a}}^{2}$, we still see a violation of inequality~\ref{eq:nonlocalitydirect} by up to 2.1 standard deviations.

An alternative method for dealing with this $N$-dependent clock shift is to post-select the data to a much narrower window in $N$, and using the measured values $(2S_{\vect{a}}/N)\si{raw}$ directly. Such a strong post-selection with $N \in [465,485]$ gives a value of $\zeta_{\vect{a}}^2 = 0.225(51)$, consistent with our previous estimate but with a lower statistical significance due to the reduced sample size of 68 points.

In a second experimental run, illustrated in Figs.~\ref{fig:stateprep}D-F, we measure $\mathcal{C}_{\vect{n}(\tau)}$ for many different axes $\vect{n}(\tau)$. For this we apply $\mathcal{R}(\kappa,\varphi_0)$ followed by a second rotation $\mathcal{R}(\vartheta(\tau),\varphi_1)$, where $\tau$ is the Rabi pulse duration, and the phase $\varphi_1$ is adjusted to maximize the contrast: this sequence ensures that the rotation is precisely around an axis perpendicular to the position of the state on the equator. Since in this second run the state has slightly shifted in phase due to  experimental drifts, the first rotation does not bring the state exactly onto the equator and we end up with a slightly different $\mathcal{C}_{\vect{n}(\tau=0)}\neq\mathcal{C}_{\vect{a}}$ even for $\tau=0$, as shown in Fig.~\ref{fig:stateprep}D. We simultaneously account for this shift and calibrate the Rabi frequency and its nonlinearity  by fitting $\mathcal{C}_{\vect{n}(\tau)}=\mathcal{C}_{\vect{b}}\sin(\tau_0 + \gamma \tau+\delta\tau^2)$ with $\{\mathcal{C}_{\vect{b}},\tau_0,\gamma,\delta\}=\{0.980(2),-0.030(9),2.464(15)\,\mathrm{ms}^{-1},-1.6(5)\times10^{-2}\,\mathrm{ms}^{-2}\}$, from which we compute $\vartheta(\tau)=\tau_0 + \gamma\tau+\delta\tau^2-\arcsin(\mathcal{C}_{\vect{a}}/\mathcal{C}_{\vect{b}})$ such that $\vect{a}\cdot\vect{n}(\tau)=\cos[\vartheta(\tau)]$.

Using the above values of $\zeta^{2}_{\vect{a}}$, $\mathcal{C}_{\vect{n}(\tau)}$, and $\vect{a}\cdot\vect{n}(\tau)$, we plot the expectation value $\mathcal{W}$ as a function of $\tau$ (see Fig.~\ref{fig:data}E). A sign of a properly calibrated angle $\vartheta(\tau)$ is seen in the four-fold symmetry of Fig.~\ref{fig:data}E.

\begin{figure}
	\begin{center}
		\includegraphics[width=0.6\textwidth]{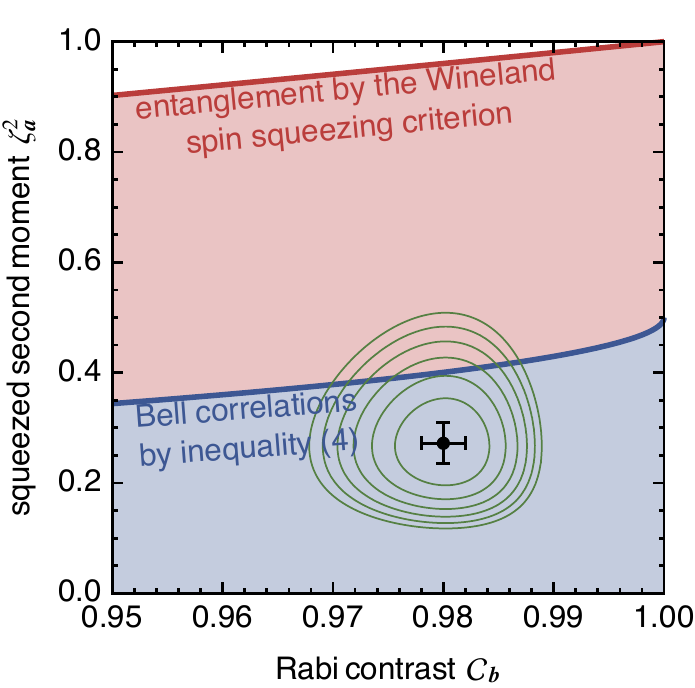}
	\caption{\label{fig:probabilityoverlaps}
		\textbf{Probability distribution describing the experimental error bars of our data point in Fig.~\ref{fig:robustviolation}.}
		The green contours contain 90\% (innermost), 99\%, \ldots, 99.9999\% (outermost) of the joint probability density for two random variables $\mathcal{C}_{\vect{b}}$ and $\zeta_{\vect{a}}^2$ as described in the text.}
	\end{center}
\end{figure}

To estimate the overlap of our data set with the blue shaded region of Fig.~\ref{fig:robustviolation}, we assume that $\mathcal{C}_{\vect{b}}$ is a random variable following a beta distribution on the interval $[-1,1]$ and $\zeta_{\vect{a}}^2$ is an independent random variable following a gamma distribution on $[0,\infty)$, with both mean values and both variances as determined experimentally. In Fig.~\ref{fig:probabilityoverlaps} their joint probability distribution is shown as green contours. A numerical integral of this joint probability distribution over the blue shaded region gives $p=0.9989$.

In the same way, we can estimate the overlap of our data set with the various $k$-producibility areas of Fig.~\ref{fig:robustviolation}. We find overlaps of $0.010$ for $k=24$ and $0.046$ for $k=29$, which allow us to rule out 24-producibility at the 1\% level and 29-producibility at the 5\%-level.

\section{Finite statistics analysis}
\label{supp:statistics}

In this section, we analyse the effect of having performed only a finite number of measurements. We  construct a quantum state that satisfies our inequality~\ref{eq:nonlocalitydirect}, yet produces a violation comparable to the one reported in Fig.~\ref{fig:data} with high probability. In future experiments aiming at  state-independent demonstrations of Bell correlations, such states must be excluded to close the statistics loophole. In order to do this, we must perform a number of measurements at least proportional to the number of particles.

Let us consider the quantum state
\begin{equation}
	\label{eq:badmix}
	\op{\rho} = (1-q)\ket{\chi}\bra{\chi} + q(\upket\upbra)^{\otimes N},
\end{equation}
where $\ket{\chi}$ is a squeezed state of $N$ spins, generated by one-axis twisting~\cite{KitagawaPRA1993} and optimized for maximally violating inequality~\ref{eq:nonlocalitydirect}; and $\upket$ is the state of one spin oriented along the squeezing axis. For $N=476$ and $\vartheta=128^{\circ}$, the values of the witness~\ref{eq:witness} evaluated on the two components of $\op{\rho}$ are
\begin{eqnarray}
	\mathcal{W}(\ket{\chi}\bra{\chi}) &=& \mathcal{W}_1\approx-0.133\nonumber\\
	\mathcal{W}\left[(\upket\upbra)^{\otimes N}\right] &=& \mathcal{W}_2\approx180.
\end{eqnarray}
By linearity of the witness, the state $\op{\rho}$ produces a value $\mathcal{W}(\op{\rho}) = (1-q)\mathcal{W}_1+q\mathcal{W}_2$. Hence, for $q\geq q^*=\frac{-\mathcal{W}_1}{\mathcal{W}_2-\mathcal{W}_1}\approx 7.38\times 10^{-4}$, this state is not considered Bell-correlated by our witness.

However, since $q^*$ is very small, whenever we make only a few measurements on $\op{\rho}$, we are most likely to sample only the squeezed state $\ket{\chi}$, which can violate our witness since $\mathcal{W}_1<0$. Taking $M$ measurements on a state with $q\ge q^*$, this happens with probability $p=(1-q)^M\le(1-q^*)^M=p^*$. In our case, our best violation $\mathcal{W}=-6.1(16)\times10^{-2}$ was achieved with $M_{\vect{n}}=10$ measurements along $\vect{n}$ and $M_{\vect{a}}=190$ along $\vect{a}$, for a total of $M=200$, hence we have $p^*\approx 0.86$. The value of $p^*$ sets a lower bound on the $p$-value of a statistical test tailored to rule out the null hypothesis ``The measured state satisfies inequality~\ref{eq:nonlocalitydirect}.'' Hence, without making any further assumptions, our finite statistics do not allow us to rule out the possibility that our state is not Bell-correlated with a confidence larger than 14\%.

A similar reasoning applies to the case of entanglement witnesses based on two-body correlators, such as those described in~\cite{SoerensenNATURE2001}. The Wineland squeezing parameter~\cite{WinelandPRA1994} of the second part of state~\ref{eq:badmix} is so different from the squeezing parameter of the first one, that a very small component $q$ is enough to make the squeezing parameter of the entire state larger than one on average. However, as long as fewer than $\ln(p)/\ln(1-q)$ measurements are performed on the system, with probability $p$ all measurements only sample the first part of the state, and thus produce statistics (including contrast and squeezing parameter) that are indistinguishable from those of a squeezed state.

In the case of our witness, the state~\ref{eq:badmix} also sets a lower bound on the $p$-value of the hypothesis ``The measured state satisfies inequality~\ref{eq:nonlocalitydirect}'' for any number of particles: for $N$ sufficiently large, one has $\mathcal{W}_1\approx -\frac14$ and $\mathcal{W}_2 \approx 0.38N$ (still for $\vartheta=128^{\circ}$). A $p$-value of 5\% is then only possible for a number of measurements $M\gtrsim4.5N$.

This conclusion can be reconciled with the observation that our inequality is violated with 3.8 standard deviations of experimental uncertainty by noting that the above state~\ref{eq:badmix} produces statistics with rare events, following a distribution that is very far from Gaussian. No mechanism is known by which states like~\ref{eq:badmix} could be produced in  our experiment; further, over many years of conducting experiments with spin-squeezed BECs, we have never observed the rare events described in the previous section, and such outliers would be easy to detect.

\section{Towards a Bell test with a many-body quantum system}
\label{supp:BellTest}

In the main text, we demonstrate that Bell correlations can be detected in a many-body system with the help of a witness observable. Unlike a Bell inequality, this witness relies on knowledge of the measurements that are performed. Still, violation of the witness inequality certifies that the measured state could be used to violate a Bell inequality. Here we discuss how one might proceed to verify this, and thus observe a Bell inequality violation with a many-body system.

In order to test a Bell inequality, one must be able to address several parties individually. In the case of Bell inequality~\ref{eq:Bellinequality} from the main text, this is required in order to observe the term $\mathcal{S}_{01}$. Ideally, all spins could be separated from each other and addressed individually. Given that the state we prepare experimentally violates inequality~\ref{eq:nonlocalitydirect}, we know that it would violate inequality~\ref{eq:Bellinequality} in this situation.

However, separating the particles into just two well-identifiable entities could in principle already be sufficient to test a Bell inequality. In a BEC, for instance, the spins could be partitioned into two groups of atoms by means of a state-independent potential. An adequate bipartite Bell inequality, different from inequality~\ref{eq:Bellinequality}, could then be tested by measuring these two groups of spins separately. Experimentally, the precision of the measurements needed to observe a bipartite Bell violation might be a limiting factor in this scenario.

A Bell inequality violation obtained in this way would be device-independent. Indeed, once a set of parties can be well identified and individually addressed, the measurements performed need not be trusted anymore. Rather, a Bell inequality violation certifies that proper measurements have been performed. Also, the outcome statistics observed in such a test could not be described in terms of pre-established agreements. 

Note that in the situation of a Bell inequality violation, one might be interested in closing the detection loophole. In this situation, post-selection of the data would not be allowed. Furthermore, one might still need to rely on the assumption that particles belonging to different parties do not communicate. This is similar to the assumption that the spins do not communicate while testing a witness. This assumption could be relaxed by separating the parties sufficiently to allow for measurements to be performed at space-like separations.

\clearpage
\bibliography{BellCorrelations}

\begin{thebibliography}{10}

\bibitem{BrunnerRMP2014}
N.~Brunner, D.~Cavalcanti, S.~Pironio, V.~Scarani, S.~Wehner, {\it Rev. Mod.
  Phys.\/} {\bf 86}, 419 (2014).

\bibitem{Bell1990}
J.~S. Bell, {\it Between Science and Technology\/}, A.~Sarlemijn, P.~Kroes,
  eds. (Elsevier, 1990), chap.~6.

\bibitem{ScaraniAPS2012}
V.~Scarani, {\it Acta Physica Slovaca\/} {\bf 62}, 347 (2012).

\bibitem{LanyonPRL2014}
B.~P. Lanyon, {\it et~al.\/}, {\it Phys. Rev. Lett.\/} {\bf 112}, 100403
  (2014).

\bibitem{EiblPRL2003}
M.~Eibl, {\it et~al.\/}, {\it Phys. Rev. Lett.\/} {\bf 90}, 200403 (2003).

\bibitem{ZhaoPRL2003}
Z.~Zhao, {\it et~al.\/}, {\it Phys. Rev. Lett.\/} {\bf 91}, 180401 (2003).

\bibitem{HofmannSCIENCE2012}
J.~Hofmann, {\it et~al.\/}, {\it Science\/} {\bf 337}, 72 (2012).

\bibitem{PfaffNATUREPHYSICS2013}
W.~Pfaff, {\it et~al.\/}, {\it Nat. Phys.\/} {\bf 9}, 29 (2013).

\bibitem{AnsmannNATURE2009}
M.~Ansmann, {\it et~al.\/}, {\it Nature\/} {\bf 461}, 504 (2009).

\bibitem{DrummondPRL1983}
P.~D. Drummond, {\it Phys. Rev. Lett.\/} {\bf 50}, 1407 (1983).

\bibitem{SvetlichnyPRD1987}
G.~Svetlichny, {\it Phys. Rev. D\/} {\bf 35}, 3066 (1987).

\bibitem{ZukowskiPRL2002}
M.~{\.Z}ukowski, {\v{C}}.~Brukner, {\it Phys. Rev. Lett.\/} {\bf 88}, 210401
  (2002).

\bibitem{AmicoRMP2008}
L.~Amico, R.~Fazio, A.~Osterloh, V.~Vedral, {\it Rev. Mod. Phys.\/} {\bf 80},
  517 (2008).

\bibitem{BlochRMP2009}
I.~Bloch, J.~Dalibard, W.~Zwerger, {\it Rev. Mod. Phys.\/} {\bf 80}, 885
  (2008).

\bibitem{GrossNATURE2010}
C.~Gross, T.~Zibold, E.~Nicklas, J.~Est{\`e}ve, M.~K. Oberthaler, {\it
  Nature\/} {\bf 464}, 1165 (2010).

\bibitem{LerouxPRL2010}
I.~D. Leroux, M.~H. Schleier-Smith, V.~Vuleti{\'c}, {\it Phys. Rev. Lett.\/}
  {\bf 104}, 250801 (2010).

\bibitem{LouchetChauvetNJP2010}
A.~Louchet-Chauvet, {\it et~al.\/}, {\it New Journal of Physics\/} {\bf 12},
  065032 (2010).

\bibitem{OckeloenPRL2013}
C.~F. Ockeloen, R.~Schmied, M.~F. Riedel, P.~Treutlein, {\it Phys. Rev.
  Lett.\/} {\bf 111}, 143001 (2013).

\bibitem{SoerensenNATURE2001}
A.~S{\o}rensen, L.-M. Duan, J.~I. Cirac, P.~Zoller, {\it Nature\/} {\bf 409},
  63 (2001).

\bibitem{SoerensenPRL2001}
A.~S. S{\o}rensen, K.~M{\o}lmer, {\it Phys. Rev. Lett.\/} {\bf 86}, 4431
  (2001).

\bibitem{RiedelNATURE2010}
M.~F. Riedel, {\it et~al.\/}, {\it Nature\/} {\bf 464}, 1170 (2010).

\bibitem{HyllusPRA2012}
P.~Hyllus, L.~Pezz{\'e}, A.~Smerzi, G.~T{\'o}th, {\it Phys. Rev. A\/} {\bf 86},
  012337 (2012).

\bibitem{TuraSCIENCE2014}
J.~Tura, {\it et~al.\/}, {\it Science\/} {\bf 344}, 1256 (2014).

\bibitem{supplementary}
Materials and methods are available as supplementary materials on \emph{Science
  Online}.

\bibitem{BancalPRL2011}
J.-D. Bancal, N.~Gisin, Y.-C. Liang, S.~Pironio, {\it Phys. Rev. Lett.\/} {\bf
  106}, 250404 (2011).

\bibitem{BoehiNATURE2009}
P.~B{\"o}hi, {\it et~al.\/}, {\it Nat. Phys.\/} {\bf 5}, 592 (2009).

\bibitem{KitagawaPRA1993}
M.~Kitagawa, M.~Ueda, {\it Phys. Rev. A\/} {\bf 47}, 5138 (1993).

\bibitem{WinelandPRA1994}
D.~J. Wineland, J.~J. Bollinger, W.~M. Itano, D.~J. Heinzen, {\it Phys. Rev.
  A\/} {\bf 50}, 67 (1994).

\bibitem{BancalPRL2009}
J.-D. Bancal, C.~Branciard, N.~Gisin, S.~Pironio, {\it Phys. Rev. Lett.\/} {\bf
  103}, 090503 (2009).

\bibitem{CurchodPRA2015}
F.~J. Curchod, N.~Gisin, Y.-C. Liang, {\it Phys. Rev. A\/} {\bf 91}, 012121
  (2015).

\bibitem{PironioNATURE2010}
S.~Pironio, {\it et~al.\/}, {\it Nature\/} {\bf 464}, 1021 (2010).

\bibitem{SchmiedNJP2011}
R.~Schmied, P.~Treutlein, {\it New J. Phys.\/} {\bf 13}, 065019 (2011).

\bibitem{OckeloenPhD2014}
C.~F. Ockeloen, Quantum metrology with a scanning probe atom interferometer,
  Ph.D. thesis, University of Basel (2014).

\end{thebibliography}
\bibliographystyle{Science}

\end{document}